\documentclass[aps,reprint,longbibliography,superscriptaddress]{revtex4-2}
\usepackage{mathrsfs}
\usepackage{amsmath,gensymb}
\usepackage{amsfonts}
\usepackage{amssymb}
\usepackage{amsthm}
\usepackage{graphicx}
\usepackage{natbib}
\usepackage{xcolor}
\usepackage{hyperref}
\usepackage{bm}
\usepackage[caption=false]{subfig}
\usepackage{verbatim}
\usepackage[normalem]{ulem}
\usepackage{soul}
\setulcolor{red}
\usepackage{xtab,afterpage}
\usepackage{lipsum}

\begin{document}

\title{%Anomalous phase shift and superconducting diode effect in Josephson junctions via $\mathrm{RET_2Si_2}$ intermetallic magnets \\
%\add{(1) 
Anomalous phase shift and superconducting diode effect in Josephson junctions via thin films of rare-earth intermetallic magnets
%}
%\\
%\add{(2) Anomalous phase shift and superconducting diode effect in Josephson junctions via $\mathrm{GdIr_2Si_2}$ thin film}
}

\author{G. A. Bobkov}
\affiliation{Moscow Institute of Physics and Technology, Dolgoprudny, 141700 Moscow region, Russia}

\author{I. A. Shvets}
\affiliation{Laboratory of Nanostructured Surfaces and Coatings, Tomsk State University, 634050 Tomsk, Russia}

\author{I. V. Bobkova}
\affiliation{Moscow Institute of Physics and Technology, Dolgoprudny, 141700 Moscow region, Russia}
\affiliation{National Research University Higher School of Economics, 101000 Moscow, Russia}

\author{A. M. Bobkov}
\affiliation{Moscow Institute of Physics and Technology, Dolgoprudny, 141700 Moscow region, Russia}

\author{S. V. Eremeev}
\affiliation{Institute of Strength Physics and Materials Science, Russian Academy of Sciences, 634055 Tomsk, Russia}

\author{E. V. Chulkov}
\affiliation{Donostia International Physics Center (DIPC), 20018 Donostia-San Sebasti\'{a}n, Basque Country, Spain}
\affiliation{Laboratory of Electronic and Spin Structure of Nanosystems, Saint Petersburg State University, 199034 Saint Petersburg, Russia}

\begin{abstract}

The superconductor/ferromagnet/superconductor (S/F/S) Josephson junctions (JJs) with an anomalous ground state phase shift $\varphi_0 \neq 0,\pi$ ($\varphi_0$-S/F/S JJs) enable the implementation of the zero-field Josephson diode effect with the possibility to control the diode efficiency and polarity. It is just as important that in this case $\varphi_0$ provides a coupling between the superconducting phase and the magnetization of the interlayer. Such $\varphi_0$-S/F/S JJs can be used for
superconducting memory and logic circuit applications. Here we present the results of theoretical calculation of the current-phase relationship (CPR), exhibiting the Josephson diode effect and $\varphi_0\neq 0,\pi$, for a JJ through a specific magnetic material. As the interlayer of the JJ we consider an ultra-thin film of intermetallic lanthanide ($Ln$)-based compound $\mathrm{GdIr_2Si_2}$. Using the  density functional theory (DFT) methods, we study the electronic structure and magnetic properties of the film. Then  the effective tight-binding Hamiltonian (TBH), demonstrating high quantitative consistency with the electronic properties obtained from DFT calculations, is constructed. The TBH is used to calculate CPR in the framework of the Bogolubov-de Gennes approach. The CPRs demonstrate a pronounced $\varphi_0$ of the order of unity and a pronounced Josephson diode effect with the diode efficiency $ \lesssim 0.3$. Moreover, the efficiency can be controlled via rotation of in-plane magnetization in the interlayer. The prospects for utilizing alternative magnetic $Ln$-based materials of the $LnT_2X_2$ family ($T$ is a transition metal and $X$ is a $p$-element from groups III-V) for the implementation in $\varphi_0$-S/F/S JJs are also discussed.

\end{abstract}

\maketitle

\section{Introduction}
\label{intr}

The Josephson effect was first discovered in 1962 \cite{Josephson1962,Josephson1964,Josephson1965} and has since sparked significant interest in both its fundamental physics and practical applications \cite{Golubov2004}. Modern applications encompass a variety of areas, including the development of sensors for detecting extremely weak magnetic fields and low-level electromagnetic radiation, the development of large-scale integrated circuits for signal processing, ultrafast digital rapid single flux quantum circuits, qubits and quantum circuits, as well as superconducting elements functioning as artificial neurons and synapses \cite{Likharev1991,Likharev2000,Soloviev2018,Ishida2021,2022_Semenov,2011_Herr,2011_Mukhanov,2012_Tanaka,2017_Soloviev,2019_Castellanos-Beltran,2022_Skryabina,2010_Crotty,2023_Schegolev,2024_Schegolev}.

The basic property of Josephson junctions (JJs) is the current-phase relation (CPR), which connects the superconducting phase difference $\varphi$ between the superconducting leads and the dissipationless electric current $I$ flowing through the junction. The minimal form of the CPR is given by $I(\varphi) = I_c \sin \varphi $, where $I_c$ is the maximal dissipationless current, which can be sustained by the JJ, called the critical current. The ground state of the JJ is reached at $\varphi=0$ (if $I_c > 0$) or $\pi$ (if $I_c <0$). The corresponding JJs are called $0$-JJ or $\pi$-JJ, respectively. In most cases, the Josephson junction is a 0-JJ. However, $\pi$-junctions have also been realized in different systems, including superconductor/ferromagnet/superconductor(S/F/S) JJs \cite{Buzdin1982,Buzdin2005,Ryazanov2001,Kontos2002,robinson2006critical}, nonequilibrium superconductor/normal metal/superconductor (S/N/S) JJs \cite{Baselmans1999,golikova2021controllable} and  others \cite{schulz2000design,jorgensen2007critical,vanDam2006,ke2019ballistic}. The $\pi$-JJs found applications in superconducting logic and quantum computers \cite{Feofanov2010,Yamashita2005,Shcherbakova2015}. 

Simultaneous breaking of time-reversal and inversion symmetries can also give rise to the anomalous ground state phase shift $\varphi_0 \neq 0, \pi$ of the JJ. The simplest sinusoidal form of the CPR is given by $I(\varphi) = I_c \sin (\varphi - \varphi_0)$ and the corresponding JJ is referred to as $\varphi_0$-JJ \cite{Bobkova_review,Shukrinov_review}. The inversion symmetry breaking leads to the noticeable Rashba-type spin splitting when the spin-orbit coupling (SOC) is strong enough, as is the case in nanowires, at surface of noble metals \cite{Ast2007}, or in some thin metallic films on substrates \cite{Mathias2010}. Another manifestation of the inversion symmetry breaking is the full spin-momentum locking, as it is realized, for example, for the 3D topological insulator (TI) surface states \cite{Burkov2010,Culcer2010,Yazyev2010,Li2014,Bihlmayer2022}. The easiest way to break the time-reversal symmetry is to apply an external magnetic field. In this case, $\varphi_0$ is a monotonic and is typically  a linear function of $h_y$ \cite{Buzdin2008,Bergeret2015,Zyuzin2016,Nashaat2019}, where $h_y$ is the component of the applied magnetic field perpendicular to the Josephson current direction and also perpendicular to the symmetry breaking axis.  
Experimentally $\varphi_0$-JJs have been realized in systems with large SOC or TIs in the interlayer region when subjected to an external magnetic field \cite{Mayer2020,Szombati2016,Assouline2019,Murani2017,Strambini2020,Reinhardt2024,Sivakumar2025}.  
Furthermore, in conjunction with higher-order current harmonics, the $\varphi_0$-JJs can manifest non-reciprocity \cite{Baumgartner2022,Pal2022,Kim2024,Reinhardt2024,Sivakumar2025}, which signifies that the magnitudes of the positive and negative critical currents are unequal, $|I_c^+| \neq |I_c^-|$. Therefore, such nonsinusoidal $\varphi_0$-JJs represent the Josephson supercurrent diodes \cite{Nadeem2023}. It should be noted here that other mechanisms of the supercurrent diode implementations also exist, see, for example Refs.~\cite{Narita2022,Narita2023,Lin2022,Golod2022,Hou2023}, Ref.~\cite{Seleznev2024} and references therein.

Another possibility to break the time-reversal symmetry is to use JJs with ferromagnetic interlayers (S/F/S JJs) instead of the applying an external field. 
In this case, the anomalous ground state phase shift $\varphi_0 \equiv \varphi_0 (m_y)$ depends on the interlayer magnetization component $m_y$ that is perpendicular to the Josephson current direction and  the symmetry breaking axis.  It is similar to the case of the applied external field, but the inherent exchange field results in the non-volatility of $\varphi_0$ at zero applied field and enables implementation of the zero-field non-reciprocal critical current with the possibility to control the polarization of supercurrent non-reciprocity by preconfiguring the magnetic state of the JJ. Such ferromagnetic Josephson diodes can be used for superconducting-phase memory and logic circuit applications 
\cite{Guarcello2020,Strambini2020,Goldobin2013}. Another advantage of the S/F/S $\varphi_0$-JJs is that their CPR provides a coupling between the electric current and the superconducting phase, on the one hand, and the magnetization, on the other hand. %\out{Based on} \add{In light of}  this, different types of the magnetization control, magnetization dynamics and magnetization switching in $\varphi_0$-S/F/S JJs have been proposed theoretically \textcolor{teal}{VARIANTS: 1.  Building on this coupling, various schemes for magnetization control, dynamics, and switching in $\varphi_0$-S/F/S JJs have been proposed theoretically. 2. 
This unique property has provided the underpinning for theoretical proposals on magnetization control, dynamics, and switching in $\varphi_0$-S/F/S JJs
%} 
\cite{Konschelle2009,Shukrinov2017,Nashaat2019,Rabinovich2019,Guarcello2020,Bobkova2020,Bobkova_review}. It is even more exciting that, besides controlling the magnetization of a single ferromagnet, the superconducting phase, being the macroscopic quantity, mediates the extremely long-range interaction between the magnetic moments of different weak links in chains of $\varphi_0$-S/F/S JJs \cite{Bobkov2022}. In turn, the long-range interaction provides a possibility to establish and control a collective magnetic behavior in such systems \cite{Bobkov2022,Bobkov2024modes,Bobkov2024many,Bobkov2024voltage}. 

In order to implement two key conditions in the interlayer region, i.e., strong SOC and exchange splitting of electron spectra at the Fermi level, it was proposed to use bilayer structures of a ferromagnetic insulator (metal)/3D TI \cite{Zyuzin2016,Nashaat2019,Bobkova2020} or a ferromagnetic insulator (metal)/material with strong SOC as interlayers. In recent years this strategy has been successfully implemented experimentally. The zero-field $\varphi_0$-JJs and polarity-switchable supercurrent rectification have been observed in proximity-magnetized Rashba-type Pt, Ta and W Josephson junctions \cite{Jeon2022,Jeon2026}. While such structures have clearly demonstrated the field-free Josephson diode effect and $\varphi_0$ behavior, they do not support genuine Josephson-current-driven magnetization dynamics. In these systems, the magnetic order in the spin–orbit coupled layer is proximity-induced by an adjacent insulating ferromagnet, and the Josephson energy remains small compared to the magnetic anisotropy energy of the entire magnet—insufficient to drive magnetization dynamics. Achieving such supercurrent-driven control therefore requires a ferromagnet with intrinsic, strong spin–orbit coupling and easy-plane anisotropy. The identification and experimental realization of such materials remains a crucial open challenge. 

Recent breakthroughs in synthesis and study of the 2D and quasi-2D ferromagnets with strong SOC along with high tunability of such systems via gating make them promising candidates for $\varphi_0$-S/F/S JJs~\cite{Gibertini2019}.
However, current theoretical understanding often relies on simplified models \cite{Krive2004,Nesterov2016,Reynoso2008,Buzdin2008,Zazunov2009,Brunetti2013,Yokoyama2014,Bergeret2015,Campagnano2015,Konschelle2015,Kuzmanovski2016,Malshukov2010,Tanaka2009,Linder2010,Dolcini2015,Zyuzin2016,Lu2015}. While these models correctly predict the symmetry properties of the anomalous phase shift, they typically operate within a quasiclassical approximation \cite{Buzdin2008,Bergeret2015,Konschelle2015,Zyuzin2016} and neglect the realistic electronic structure of the ferromagnet. To date, no theoretical studies have fully accounted for key material-specific characteristics, such as strong exchange splitting, SOC comparable to the band width (of the order of eV), or the presence of multiple electronic bands at the Fermi level. Therefore, to advance this field, a targeted search for suitable materials, guided by computational predictions that incorporate these realistic features, is essential.

\begin{figure}[t]
\includegraphics[width=\columnwidth]{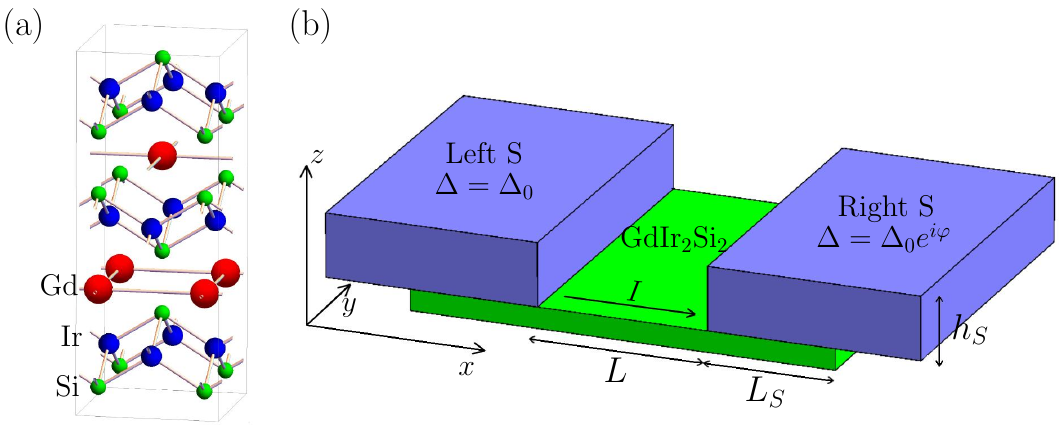}
\caption{(a) Translationally invariant along the $(x,y)$-plane minimal cell of the $\mathrm{GdIr_2Si_2}$ symmetric slab of the $I4/mmm$-phase with Si-termination, which is  used as an interlayer of the JJ. (b) Sketch of the planar JJ via the $\mathrm{GdIr_2Si_2}$ interlayer. The length of the magnetic weak link is $L$, the length of the  magnet covered by the superconductor is $L_s$, the thickness of the superconducting leads is $h_S$. } 
 \label{fig:system}
\end{figure}

The aim of this work is to implement the described strategy. Using a combination of density functional theory (DFT) methods and Josephson current calculations in the formalism of the Bogoliubov-de Gennes equations, we investigate the possibility of realizing a $\varphi_0$-JJ through thin films of the rare-earth intermetallic magnet $\mathrm{GdIr_2Si_2}$, which belongs to the $LnT_2X_2$ family ($Ln$ -- lanthanide, $T$ -- transition metal). This family represents a beautiful playground for systematic
studies of the emergence of the Rashba effect and its
properties because the strength of the SOC can be tuned by choosing suitable transition metal atoms. It increases by exchanging transition metals from $3d$ to $4d$ and further to $5d$ \cite{Schulz2021,Schulz2019,Generalov2018,Usachev2020,Nechaev2018}. The sketch of the JJ and the crystal structure of the weak link material is presented in Fig.~\ref{fig:system}. In general, we obtained non-sinusoidal CPR, which can be well-approximated by the first and second harmonics contributions with their own anomalous phase shifts $\varphi_{0,1(2)}$ depending on $m_y$ in a complex, highly non-monotonic manner.  Due to the strong exchange splitting in combination with strong SOC the  electronic structure near the Fermi level is highly anisotropic. For this reason, the critical currents, $I_{c1}$ and $I_{c2}$, 
and superconducting Josephson diode efficiency also manifest strong anisotropy with respect to the mutual orientation of the Josephson current and the in-plane magnetization direction. On the other hand, we predict that the considered  $\mathrm{GdIr_2Si_2}$ thin film is an easy-plane magnet. Thus, the value of the anomalous phase shift and the diode efficiency can be adjusted via small rotations of the in-plane magnetization. These findings hold significant importance for future applications in spintronics, facilitating both the control of magnetization through the Josephson current and phase, and the regulation of the polarity and amplitude of the Josephson diode effect through the manipulation of magnetization direction.

The paper is organized as follows. In Sec.~\ref{sec:material} we describe the material compound and present results of DFT calculations of its electronic states dispersion and spin structure. The magnetic properties of the $\mathrm{GdIr_2Si_2}$ thin films are also discussed. In Sec.~\ref{sec:hamiltonian} we construct an effective tight-binding Hamiltonian for description of the electronic transport in $\mathrm{GdIr_2Si_2}$. In Sec.~\ref{sec:model} the model of the superconductor/$\mathrm{GdIr_2Si_2}$/superconductor JJ and the method used for calculation of the CPR are described. In Sec.~\ref{sec:cpr} we present results of calculations of CPRs including the discussion of the anomalous phase shift and the Josephson diode effect. Sec.~\ref{sec:discussion} is devoted to a qualitative discussion of other candidate materials and in Sec.~\ref{sec:conclusions} our conclusions are summarized. The Appendix presents an analysis of the sensitivity of the key Josephson characteristics to variations in the superconducting lead parameters and temperature.

\section{Compound information. Dispersion and spin structure of electronic states of the $\mathrm{GdIr_2Si_2}$ thin films}
\label{sec:material}

Electronic structure calculations were carried out within the DFT, using the projector augmented-wave (PAW) method \cite{Blochl.prb1994} as implemented in the VASP package  \cite{vasp1,vasp2}. The exchange-correlation energy was treated using the generalized gradient approximation (GGA) \cite{Perdew.prl1996}. The standard Gd potential, in which the 4$f$ electrons are treated as valence states, was used for spin-polarized calculations of magnetic phases. To correctly describe the highly correlated Gd-4$f$ and Ir-5$d$ electrons, we included the correlation effects within the GGA+$U$ method~\cite{Dudarev}. The effective Hubbard parameter $U^\ast=U-J$ was chosen to be 6 eV for Gd and 3.5 eV for Ir atom~\cite{tesch2022hubbard}, respectively. For an accurate description of interatomic distances during crystal structure optimization, van der Waals corrections with Becke–Johnson damping (DFT-D3) were applied \cite{Grimme2011}. The experimental lattice constants ($a = 4.042$, $c = 9.986$~\AA~were adopted from Ref.~\cite{Kliemt2020}. The atomic positions were optimized by minimizing forces acting on the atoms. 

\begin{figure}[t]
\includegraphics[width=\columnwidth]{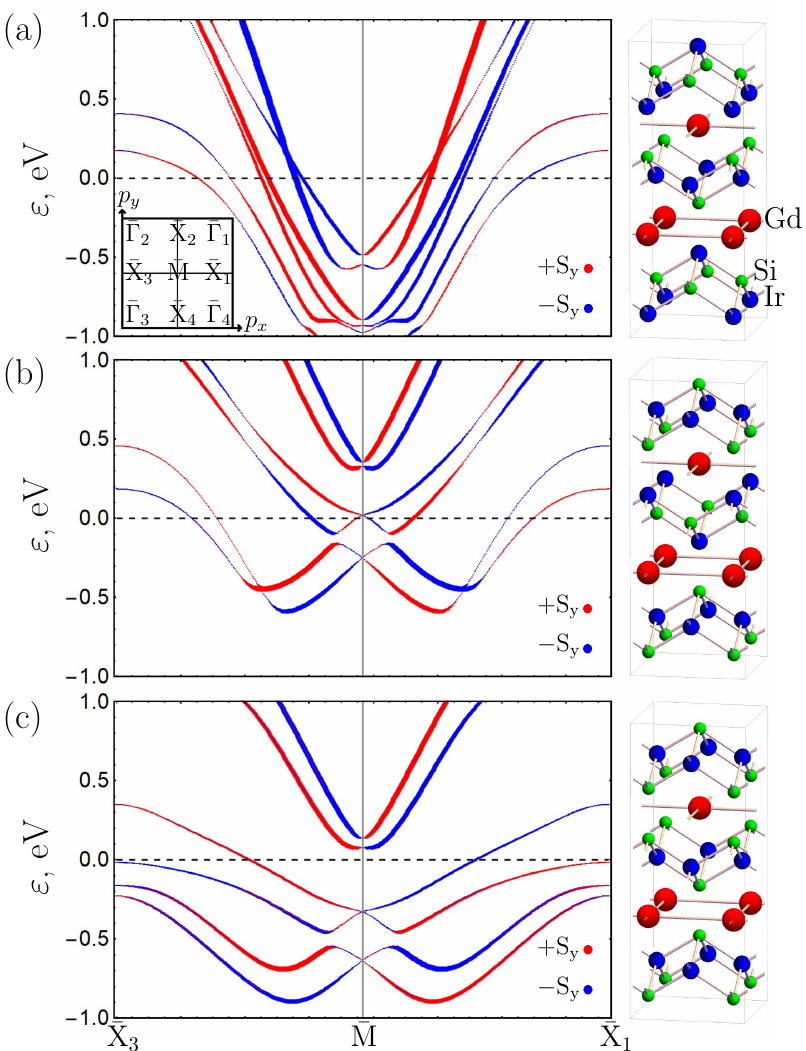}
\caption{DFT-calculated electron spectra along the $\bar{\rm X}_3 \bar{\rm M} \bar{\rm X}_1$ direction [see inset in (a)] and corresponding atomic structure for the $\mathrm{GdIr_2Si_2}$ slab (shown on the right). (a) P-phase  (Ir-termination), (b) P-phase (Si-termination), (c) I-phase (Si-termination). In all cases, magnetic moments of gadolinium atoms are ordered ferromagnetically in each of the layers with magnetization lying in-plane (along the $x$-axis), while the interlayer ordering is antiferromagnetic. Spin expectation value $S_y$ is shown by color and thickness. The spin quantization axis is chosen perpendicular to the magnetization direction to clearly resolve the spin-orbit splitting of the bands without contamination from the Zeeman contribution.} 
 \label{fig:different_phases}
\end{figure}

The electronic structure of the GdIr$_2$Si$_2$ thin film, which we suggest using as a weak link of the JJ, was studied within the framework of the slab model. The symmetric slab is constructed in such a way that its top and bottom surfaces have a silicon (iridium) termination and the slab thickness amounts to 14 atomic layers. We analyze different possible crystal configurations, including low-temperature (I4/mmm) and high-temperature (P4/nmm) structures, hereinafter referred to as the I- and P-phases, respectively (Fig.~\ref{fig:different_phases}). For the P-phase, thin films with both iridium [Fig.~\ref{fig:different_phases}(a)] and silicon [Fig.~\ref{fig:different_phases}(b)] terminations are possible, whereas in the highly symmetric I-phase only silicon termination is realized [Fig.~\ref{fig:different_phases}(c)].

{\it Specific features of the band structure.} For all three possible configurations of the slab, its band structure  has three electronic bands near the Fermi level:  %\hl{In some cases, all 
three bands intersect the Fermi level in the Ir-terminated P-phase film [see Fig.~\ref{fig:different_phases}(a)], while in other cases, only two do [Fig.~\ref{fig:different_phases}(b)-(c)]. Each band has a Rashba-type spin-orbit splitting, which manifests itself as the momentum-antisymmetric spin polarization $\bm s$ of the spectra $\varepsilon(\bm p, \bm s)=\varepsilon(-\bm p, -\bm s)$.
%, see Fig.~\ref{fig:different_phases}. 
In addition, if the material is in the magnetic state, each band also has exchange splitting. In the spectra, it is clearly visible as a splitting in high symmetry points  $\bar{\rm M}$, $\bar\Gamma$, and $\bar{\rm X}$, because the spin-orbit splitting in such points should be zero. Also, the presence of the exchange splitting violates the full momentum antisymmetry of the electron spin polarization $\varepsilon(\bm p, \bm s)=\varepsilon(-\bm p, -\bm s)$, as seen from Fig. \ref{fig:specific_phase}. This condition is only preserved in the magnetic state if the spin quantization axis is chosen perpendicular to the magnetization direction. This case is demonstrated in Fig.~\ref{fig:different_phases} to clearly show the spin-orbit splitting of the spectra.

\begin{center}
\begin{table}
\begin{tabular}{|c|c|c|c|}
\hline
~ & P4/nmm(Si) & P4/nmm(Ir) & I4/mmm(Si) \\
\hline
AF(x) & -51.32769 & -50.42999 & -51.29959 \\
\hline
AF(z) & -51.32706 & -50.42967 & -51.29977 \\
\hline
F(x) & -51.32519 & -50.43464 & -51.30899 \\
\hline
F(z) & -51.32659 & -50.43269 & -51.30930 \\
\hline
PM & -41.72508 & -40.82911 & -41.6893 \\
\hline
\end{tabular}
\caption{\label{tab:magnetic_ordering}DFT-calculated energy per magnetic atom for the three configurations of $\mathrm{GdIr_2Si_2}$ slab and five possible magnetic orderings: antiferro- and ferromagnetic with magnetization lying in-plane ($x$) and out-of-plane ($z$), and paramagnetic state (PM). Energy is given in eV/m.a.}
\end{table} 
\end{center}

%\noindent
%\hl{===========================}

{\it Magnetic ordering.} Considered $\mathrm{GdIr_2Si_2}$ symmetric slab has two layers of gadolinium atoms. The magnetic moments of both layers are ordered ferromagnetically within each of the layers. The interlayer ordering can be either ferromagnetic (F) or antiferromagnetic (AF). The magnetic moment of the layers can have in-plane (IP) or out-of-plane (OOP) orientation. Magnetocrystalline anisotropy constants can be found from the data collected in Tab. \ref{tab:magnetic_ordering}, where DFT-calculated energies per magnetic atom (m.a.) of gadolinium for all possible magnetic ordering are presented. For ferromagnetic state we should add energy of stray magnetic fields:
\begin{align}
    E^{\rm st}=-\sum_{ij} \frac{\bm\mu_j(3(\bm \mu_i \bm r_i)\bm r_i-\bm \mu_i |\bm r_i|^2)}{2|\bm r_i-\bm r_j|^5}
\end{align}
where $\bm r_{i}$ is a radius vector and $\bm \mu_i$ is a magnetic moment of gadolinium atoms, and $|\bm \mu_i|=7~\mu_\mathrm{B}$. In our cases, we have obtained $E^{\rm st}_{\rm F(z)}-E^{\rm st}_{\rm F(x)}=+1.31$~meV/m.a. (which in all cases exceeds the magnetocrystalline anisotropy). As a result, we can find a favorable magnetic state:

P-phase(Si): IP magnetization, AF interlayer ordering;

P-phase(Ir): IP magnetization, F interlayer ordering;

I-phase(Si): IP magnetization, F interlayer ordering.

Ir-termination has significantly higher energy than Si-termination. Of the two silicon terminations, we chose the I-phase film for further consideration since it has much higher T$_N$ as compared to the P-phase \cite{Kliemt2020}.
%, for further consideration, which will be studied here as the weak link of the JJ. \comm{(we need motivation for choosing I-phase among three variants of thin films. It is known that in bulk crystals the antiferromagnetic order is established at T$_N$ = 86 K for the I-phase compound, whereas the compound with the CaBe$_2$Ge$_2$ structure (P-type) orders antiferromagnetically at T$_N$ = 12.5 K \cite{Kliemt2020}. We can use this reason, since in thin film T$_N$ would be even smaller.)} 
In this case magnetocrystalline anisotropy $E_{\rm {F(z)}}^{\rm{cr}}-E_{\rm {F(x)}}^{\rm{cr}}=-0.31$~meV/m.a., where $E_{\rm {F(z)}}^{\rm{cr}}$ and $E_{\rm {F(x)}}^{\rm{cr}}$ are energies per unit cell of the slabs with the OOP and IP ferromagnetic interlayer ordering, respectively. Energy of stray magnetic fields $E_{\rm {F(z)}}^{\rm{st}}-E_{\rm {F(x)}}^{\rm{st}}=+1.31$~meV/m.a. is higher due to the very large magnetic moment of gadolinium and makes the IP orientation energetically more favorable with the total anisotropy $E_{\rm OOP}-E_{\rm IP}= [E_{\rm {F(z)}}^{\rm{cr}}-E_{\rm {F(x)}}^{\rm{cr}}]+ [E_{\rm {F(z)}}^{\rm{st}}-E_{\rm {F(x)}}^{\rm{st}}] =  +0.99$~meV/m.a. 

Such a high magnetic anisotropy constant ensures that the magnetic moment will always have IP orientation. Therefore, in the following, we can describe the magnetic orientation via the single
parameter $m_y\in [-1,1]$, which is the $y$-axis projection of the unit vector $\bm m$ along the magnetization $\bm M=|\bm M|(\sqrt{1-m_y^2}, m_y, 0)$.

\section{Tight-binding model for description of electronic subsystem of $\mathrm{GdIr_2Si_2}$}
\label{sec:hamiltonian}

\begin{figure}[t]
\includegraphics[width=\columnwidth]{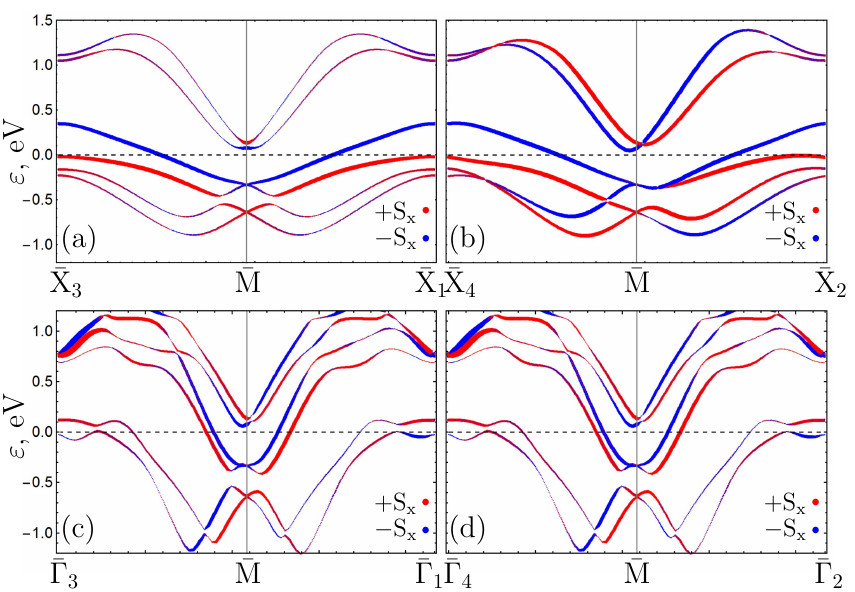}
\caption{DFT-calculated electronic spectra for the I-phase $\mathrm{GdIr_2Si_2}$ film, used in calculations of the Josephson characteristics [for crystal structure see Fig.~\ref{fig:different_phases}(c)]. The magnetization is aligned with the $x$-axis.} 
 \label{fig:specific_phase}
\end{figure}

From Fig.~\ref{fig:specific_phase} one can conclude that the Fermi level is crossed by two electron bands, each of which has a spin-orbit splitting. Let us designate the lower energy band as $I$ and the upper one as $II$. The band $II$ has a cubic Rashba-type spin-orbit splitting, while the band $I$ has a linear Rashba-type spin-orbit splitting. 
It is important to note that, in fact, each of the electronic bands is twofold degenerate, since in the limit of a symmetric ultrathin film, each of these states belongs to one of two surfaces of the film. Due to the negligible splitting of these states, we can conclude that their interaction is very small and can be disregarded. For this reason, in what follows, we will describe one set of states, localized at the upper surface of the slab and assume that they will solely participate in the transfer of current. The bottom surface has a negligible effect on the total current amplitude due to its extremely weak coupling to the superconducting leads.

We use the following effective tight-binding Hamiltonian (TBH) to describe non-superconducting state of the ${\rm GdIr_2Si_2}$ slab:
\begin{align}
\hat H=-\sum_{\bm r,\bm r';\nu,\nu';\sigma,\sigma'} \hat{c}_{\bm r\nu\sigma}^{\dagger}\hat t_{\bm r \bm r';\sigma\sigma'}^{(\nu, \nu')}\hat{c}_{\bm r'\nu'\sigma'},
\label{eq:TBH}
\end{align}
where $\hat c_{\bm r\nu\sigma}$ is the electron annihilation operator, $\bm r$ is the radius vector of the unit cell, $\nu=I,II$  is the index of the corresponding conduction band, $\sigma=\uparrow,\downarrow$ is the spin projection on the $z$-axis. The main goal is to select a minimal set of non-zero hopping matrix elements $\hat t_{\bm r \bm r';\sigma\sigma'}^{(\nu, \nu')}$ such that the TBH band structure $\varepsilon_n^{\rm TBH}(\bm p)$ matches well with the DFT band structure $\varepsilon_n^{\rm DFT}(\bm p)$. Here $n$ is the electron branch number. As stated earlier, we consider two bands, each of which has spin splitting. Accordingly, there are exactly 4 energy branches. 

Hopping element between fixed unit cells with radius vectors $\bm r$ and $\bm r'$, $\hat t_{\bm r \bm r';\sigma\sigma'}^{(\nu, \nu')}$ is a $4\times 4$ matrix in spin and band spaces. Introducing Pauli matrices  $\sigma_\alpha$ in spin space and  $\rho_\alpha$ in band space, $\alpha=0,x,y,z$, we can write:
\begin{align}
    \hat t_{\bm r \bm r';\sigma\sigma'}^{(\nu, \nu')}=\sigma_0 \left(
\begin{array}{cc}
t^{(I,I)}_{\bm r \bm r'} &  t^{(I,II)}_{\bm r \bm r'} \\
t^{(I,II)}_{\bm r \bm r'} &  t^{(II,II)}_{\bm r \bm r'}  
\end{array}
\right)_\rho+ \nonumber \\
\bm \sigma\left(
\begin{array}{cc}
\bm V^{(I,I)}_{\bm r \bm r'} &  \bm V^{(I,II)}_{\bm r \bm r'} \\
\bm V^{(I,II)*}_{\bm r' \bm r} &  \bm V^{(II,II)}_{\bm r \bm r'}  
\end{array}
\right)_\rho ,
\label{eq:t_first}
\end{align}
where $\bm \sigma=(\sigma_x,\sigma_y,\sigma_z)$ is a vector of Pauli matrices. The first term in Eq.~(\ref{eq:t_first}) is spin independent part (kinetic energy). The second term is spin dependent and includes exchange field and spin-orbit interaction:
\begin{align}
    \bm V_{\bm r \bm r'}^{(\nu, \nu')}=\left(
\begin{array}{c}
h_M^{(\nu, \nu')} m_x \\
h_M^{(\nu, \nu')} m_y \\
h_M^{(\nu, \nu')} m_z+h_z^{(\nu, \nu')}
\end{array}
\right) - \nonumber \\
i \alpha^{(\nu, \nu')} \left(
\begin{array}{c}
-\sin (N^{(\nu, \nu')}\theta_{\bm r'-\bm r})  \\
\cos (N^{(\nu, \nu')}\theta_{\bm r'-\bm r}) \\
0
\end{array}
\right) ,
\end{align}
where $\alpha^{(\nu, \nu')}$ is constant of intraband ($\nu = \nu'$) or interband ($\nu \neq \nu'$) spin-orbit coupling, $\theta_{\bm r'-\bm r}$ is an angle between vector $\bm r'- \bm r$ and $x$-axis, $N^{(\nu, \nu')}$ is an order of Rashba-type spin-orbit coupling. As it was stated earlier $N^{(I,I)}=1$, $N^{(II,II)}=3$. However, the value of $N^{(I,II)}$ is unknown. The only way is to try to fit the DFT band structure with different $N^{(I,II)}$ and choose the appropriate one. $N^{(I,II)}=1$ provides very good agreement between DFT and TBH bands. 

At first glance, the exchange field should be co-directed with the normalized magnetization $\bm m$. But an accurate analysis of the conduction bands, calculated within DFT, shows that part of the exchange splitting corresponds to the spin quantization axis, which does not rotate following the magnetization orientation and is even preserved in the paramagnetic phase. This is probably due to the fact that we are considering only two conduction bands, and this spin splitting arises from their interaction with other bands located far from the Fermi level. To correctly describe this spin splitting, we add an additional parameter $h_z^{(\nu, \nu')}$ to the exchange term, which is always directed along the $z$-axis and does not depend on the direction of $\bm M$. 
As indicated above, there is no difference between ferromagnetic and antiferromagnetic coupling of two Gd layers, since the states of upper and bottom sides of the film essentially do not interact.

\begin{figure}[t]
\includegraphics[width=0.85\columnwidth]{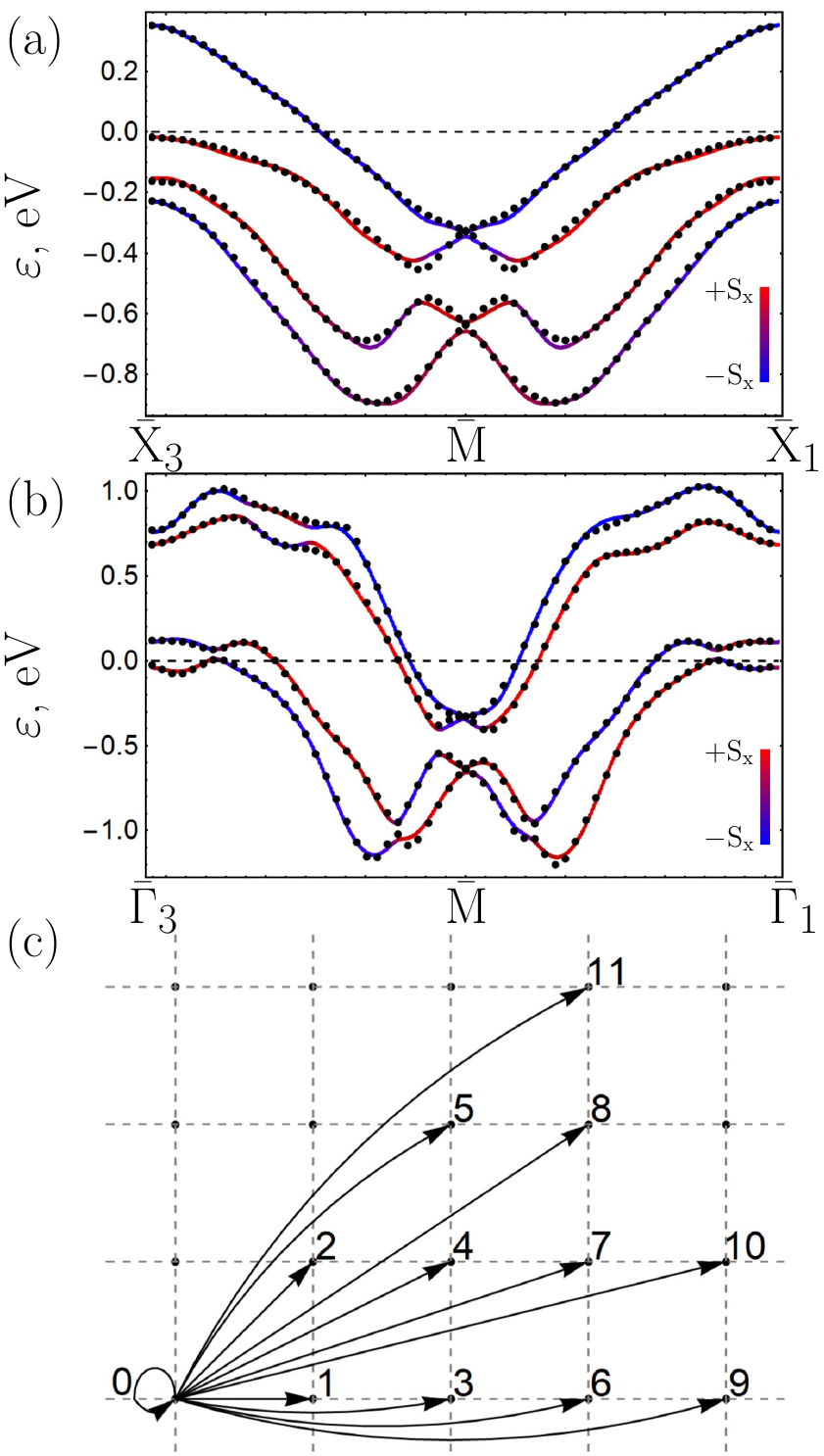}
\caption{(a) TBH electronic spectra  $\varepsilon_n^{\rm{TBH}}(\bm p)$ (lines) and DFT electronic spectra  $\varepsilon_n^{\rm{DFT}}(\bm p)$ (dots) calculated along  $\bar{\rm X}_3 \bar{\rm M} \bar{\rm X}_1$ momentum direction exploiting the hopping amplitudes presented in Tab.~\ref{tab:hopping_par}. The $x$ spin component of the $\varepsilon_n^{\rm{TBH}}(\bm p)$ states is shown by color. The magnetization is aligned with the $x$-axis, $\bm m = \bm e_x$. (b) The same for $\bar\Gamma_3 \bar{\rm M} \bar\Gamma_1$ momentum direction. (c) Sketch of all 12 hops (numbered 0--11) used for construction of the tight-binding Hamiltonian.} 
 \label{fig:fits}
\end{figure}

We use 12 nearest hopping elements (including on-site), 
the smaller number gives a poor fit to the DFT band structure.  The involved hops are schematically shown in Fig.~\ref{fig:fits}(c). The fitting was performed by minimizing the mean square error:
\begin{align}
    {\rm MSR}=\sum_{l; k; n} |\varepsilon_n^{{\rm DFT}}(\bm p_k^l,\bm m_l)-\varepsilon_n^{{\rm TBH}}(\bm p_k^l,\bm m_l)|^2,
\end{align}
where $l=1,...,8$ is a set of four 
high-symmetry directions in the surface Brillouin zone used for fitting, including ${\rm \bar X_3\bar M\bar X_1,\bar \Gamma_3\bar M\bar \Gamma_1,\bar X_4\bar M \bar X_2,\bar \Gamma_4\bar M\bar \Gamma_2}$ momentum directions, and two orientations of magnetization along the $x$- and $z$-axes. $\bm p_k$ is a set of all available points for each DFT-calculated band dispersions. Minimization ${\rm MSR}$ was performed using the gradient descent method with all 141 parameters of the TBH Hamiltonian. The final amplitudes are given in Tab.~\ref{tab:hopping_par}. The TBH electronic spectra $\varepsilon_n^{\rm{TBH}}(\bm p)$, calculated by exploiting the hopping amplitudes presented in Tab.~\ref{tab:hopping_par}, are plotted in Fig.~\ref{fig:fits}(a)-(b). For comparison, the DFT-calculated spectra $\varepsilon_n^{\rm{DFT}}(\bm p)$ are plotted by dots on the same figures. The suggested Hamiltonian provides good quantitative
agreement with the DFT spectra and the corresponding electronic spin structure for all studied momentum directions in the Brillouin zone.

\begin{center}
\begin{table*}
\begin{tabular}{|c|c|c|c|c|c|c|c|c|c|c|c|c|}
\hline
\# & $t^{(I),(I)}$ & $t^{(II),(II)}$ & $t^{(I),(II)}$ & $\alpha^{(I),(I)}$ & $\alpha^{(II),(II)}$ & $\alpha^{(I),(II)}$ & $h_M^{(I),(I)}$ & $h_M^{(II),(II)}$ & $h_M^{(I),(II)}$ & $h_z^{(I),(I)}$ & $h_z^{(II),(II)}$ & $h_z^{(I),(II)}$ \\
\hline
0 & 142.0 & -113.2 & -0.3 & - & - & - & 15.6 & 117.1 & 1.6 & 5.9 & -1.2 & 3.3 \\
\hline
1 & 148.6 & 73.0 & 5.8 & -47.7 & -22.8 & 1.7 & -1.6 & -0.8 & -0.3 & 15.3 & -0.8 & 0.5 \\
\hline
2 & 111.5 & -76.8 & 6.8 & 51.2 & -36.9 & -3.0 & -0.5 & -10.4 & -0.7 & 7.9 & -1.9 & -0.2 \\
\hline
3 & -21.2 & 76.7 & 0.7 & -42.5 & 3.4 & 7.0 & -0.2 & 6.7 & 1.6 & -0.4 & 2.7 & 1.3 \\
\hline
4 & -48.6 & 75.5 & 2.6 & 13.7 & 9.6 & 2.6 & 7.0 & 0.6 & -2.1 & -6.9 & 1.9 & -10.2 \\
\hline
5 & 8.7 & -104.4 & -6.4 & 10.8 & 4.8 & -8.3 & -1.1 & -7.4 & 1.4 & -7.7 & 1.0 & -4.0 \\
\hline
6 & 26.3 & -49.0 & -3.6 & 4.2 & 8.5 & 19.0 & -7.9 & -6.1 & 2.5 & -3.7 & -0.5 & 10.2 \\
\hline
7 & -48.7 & 39.8 & 2.4 & -3.3 & -13.6 & 15.8 & 3.7 & -1.6 & -2.5 & -9.5 & 1.3 & 6.9 \\
\hline
8 & 20.4 & -5.5 & -0.7 & 12.8 & -3.3 & -9.1 & -1.6 & 5.3 & 1.4 & 0.4 & 1.1 & -2.7 \\
\hline
9 & 22.4 & 27.3 & -8.4 & 8.9 & -6.9 & 12.5 & -1.6 & 0.5 & 1.0 & 0.1 & 1.7 & 12.9 \\
\hline
10 & -26.5 & -6.1 & -2.5 & 5.2 & 3.8 & 14.1 & 2.1 & 0.0 & 0.3 & 3.8 & 1.1 & 12.7 \\
\hline
11 & -3.8 & -2.2 & 3.1 & -1.1 & -10.9 & -14.2 & 1.8 & -1.9 & -2.4 & 5.6 & -0.3 & -1.7 \\
\hline
\end{tabular}

\caption{\label{tab:hopping_par}Parameters of the two-band tight-binding model (Eq.~\ref{eq:TBH}) fitted to the DFT-calculated electron spectra. All amplitudes are given in meV.}
\end{table*} 
\end{center}

Let us comment on the obtained number of hopping elements (141). At first glance, it seems that this number is huge and clearly excessive. It is produced by two conduction bands [($I$)-($I$), ($II$)-($II$), ($I$)-($II$)], 4 types of interaction (kinetic energy, spin-orbit interaction and exchange field of two types) and twelve nearest hops. It is worth noting that these 12 ``neighbours” merely provide an efficient means to construct a Hamiltonian that accurately reproduces the electronic structure near the Fermi level obtained from DFT calculations, while also enabling the solution of inhomogeneous superconducting problems for the material in question using standard methods of superconductivity theory, such as the Bogoliubov–de Gennes and Gor’kov equations. In the classical method of localized Wannier functions, the number of necessary nearest hops turns out to be significantly smaller, but such a procedure must take into account many electron bands at once (ideally all of them, although not necessarily). Due to the presence of only two bands in our consideration, the corresponding wave functions become less localized -- as a result, there is no excess of parameters here. Again, unlike direct wannierization, which assumes taking into account all possible terms in the Hamiltonian, we, using the physical properties of a specific band structure, have restricted ourselves to the necessary minimum. To summarize, the number of parameters used is actually very small for a successful description of such a complex band structure. Since it is clearly seen that the wave functions are not strongly localized, the $\rm  k\cdot \rm  p$  method would probably require fewer parameters, however, it cannot be further used for finite-size structure calculations using the BdG technique.

\section{S/$\mathrm{GdIr_2Si_2}$/S Josephson junction: model and method}
\label{sec:model}

To calculate the Josephson current, let us consider the system shown in Fig.~\ref{fig:system}(b). It represents a planar Josephson junction made of superconducting leads and a ${\rm GdIr_2Si_2}$ weak link (hereinafter referred to as the F layer). Geometric parameters $L_S=3a$ (total length of superconducting lead is $9a$), $h_S=3a$ and different $L$ were used. The system is modeled by the following tight-binding Hamiltonian on a square lattice:
\begin{align}
\hat{H} =  -  \sum\limits_{\bm i \bm j, \sigma\sigma'} t_{\bm i \bm j;\sigma\sigma'} c^{\dagger}_{\bm i \sigma} c_{\bm j \sigma'} + 
\sum\limits_{\bm i} (\Delta_{\bm i}  c^{\dagger}_{\bm i \uparrow} c^{\dagger}_{\bm i \downarrow} + h.c.) 
\label{eq:hamiltonian}
\end{align}
Here $c_{\bm i,\sigma}$ is an annihilation operator for electrons at site $\bm r$ and we denote $\bm i\equiv \bm r$ for sites in S and $\bm i\equiv (\bm r, \bm \nu)$ for sites in F. $\sigma = \uparrow, \downarrow$ is a spin of electrons. Detailed expressions for $t_{\bm i \bm j;\sigma\sigma'}$ in the F layer were discussed in Sec.~\ref{sec:hamiltonian}. In this work, we do not focus on the detailed properties of the superconductor; it is modeled using only nearest-neighbor hopping with an amplitude of $t_S = 100$ meV. The superconducting order parameter $\Delta_{\bm i}$ is set to $\Delta_0 = 10$ meV for the S layers, and the hopping element between the superconductor and the weak link is taken as $t_{SF} = 50$ meV for the nearest-neighbor coupling to both electron bands of $\mathrm{GdIr_2Si_2}$. These values of $\Delta_0$ and $t_{SF}$ are somewhat higher than typical experimental estimates due to computational constraints. For instance, in conventional superconductors such as Nb, $\Delta_0 \approx 1.5$ meV, while typical values for interlayer hopping between layered materials are on the order of $t_{SF} \sim 30$ meV \cite{Hsu2022, Bobkov2024_hybridization}. Our approach is justified because the essential features of the current-phase relation—namely the anomalous phase shift and the functional dependence of the critical current on $m_y$—are largely insensitive to the specific superconducting material and are predominantly governed by the properties of the weak link.

In the Appendix, we present a study of the sensitivity of $\varphi_0(m_y)$ and $I_c(m_y)$ to variations in $\Delta_0$ and $t_{SF}$, including a limited set of results with more realistic values of these parameters. The results demonstrate that, while quantitative details do exhibit some dependence on these parameters, all key qualitative features are robust and remain determined solely by the intrinsic material properties of the interlayer.

The only characteristic of the junction that exhibits a significant dependence on $\Delta_0$ and $t_{SF}$ is the magnitude of the diode effect. This behavior is well documented in the literature \cite{Baumgartner2022,Pal2022,Kim2024,Reinhardt2024,Sivakumar2025,Nadeem2023,Jeon2022,Jeon2026} and represents a physically natural consequence of the fact that the diode effect requires the presence of the second or higher harmonics in the Josephson current. The relative amplitude of these harmonics inevitably diminishes as superconductivity weakens, regardless of the specific system under consideration. Nonetheless, as shown in the Appendix, the diode effect retains appreciable strength for experimentally relevant values of the superconducting and interface parameters. Importantly, the key functionality—tunability via small variations of $m_y$—remains fully intact, highlighting the practical relevance of our predictions for realistic experimental conditions.

All calculations were performed at a temperature $T = 0.1\Delta_0$. In the Appendix, we also  present the data illustrating the sensitivity of the anomalous phase shift $\varphi_0(m_y)$, critical current $I_c(m_y)$, and diode efficiency $\eta$ to variations in temperature. The results demonstrate that, similar to the influence of the superconducting environment, temperature variations do not qualitatively affect $\varphi_0(m_y)$ or the functional form of $I_c(m_y)$. The physical origins of the observed quantitative dependence of these quantities on the superconducting environment and temperature are discussed in detail in the Appendix. 

In general, the superconducting order parameter should be calculated self-consistently as $\Delta_{\bm i} = \lambda \langle c_{\bm i \downarrow} c_{\bm i \uparrow} \rangle$, where $\lambda$ is the pairing constant. If the height of the leads is less than the superconducting coherence length $\xi_S$, the superconducting order parameter is approximately homogeneous in the leads, although its amplitude can be suppressed due to proximity to the ferromagnetic layer. As it was already stated, we are not interested in the exact value of the general amplitude constant of the critical current. For this reason, here the order parameter in the leads is not calculated self-consistently.

Calculation of the Josephson current through the ${\rm GdIr_2Si_2}$ thin film was performed within the framework of Bogoliubov-de Gennes (BdG) equations. We diagonalize the Hamiltonian (\ref{eq:hamiltonian}) by the Bogoliubov transformation:
\begin{align}
c_{\bm i\sigma}=\sum\limits_n u_{n\sigma}^{\bm i}\hat b_n+v^{\bm i *}_{n\sigma}\hat b_n^\dagger, 
\label{bogolubov}
\end{align}
Then the resulting BdG equations take the form:
\begin{align}
 \sigma \Delta_{\bm i} v^{ \bm i}_{n,-\sigma} - \sum\limits_{ \bm i',\sigma'} t_{\bm i \bm i';\sigma\sigma'} u^{\bm i'}_{ n, \sigma'} & = \varepsilon_n u_{n,\sigma}^{\bm i} \nonumber \\  
\sigma \Delta_{ \bm i}^* u^{\bm i}_{n,-\sigma} - \sum\limits_{ \bm i',\sigma'} (t_{\bm i \bm i';\sigma\sigma'})^* v^{\bm i'}_{ n, \sigma'} & = -\varepsilon_n v_{n,\sigma}^{\bm i}, 
\label{bdg}
\end{align}

The current between sites $\bm i$ and $\bm i'$ can be calculated via the solutions of the BdG equation as follows:
\begin{align}
    \bm j_{\bm i \to\bm i'}=e\sum_{n,\sigma,\sigma'} i  \left[(t_{\bm i \bm i';\sigma\sigma'}u_{n\sigma}^{ \bm i} u_{n\sigma'}^{\bm i'*}-t^*_{\bm i \bm i';\sigma\sigma'}u_{n\sigma'}^{ \bm i'} u_{n\sigma}^{\bm i*})f_n \right. \nonumber \\  
    \left.+(t^*_{\bm i \bm i';\sigma\sigma'}v_{n\sigma}^{ \bm i*} v_{n\sigma'}^{\bm i'}-t_{\bm i \bm i';\sigma\sigma'}v_{n\sigma'}^{ \bm i'*} v_{n\sigma}^{\bm i})(1-f_n)\right],
\end{align}
where $f_n=\frac{1}{e^{\varepsilon_n/T}+1}$ is a Fermi–Dirac distribution. The total linear current density (per unit length along the $y$-axis) can be obtained via summation of all $\bm j_{\bm i \to\bm i'}$ that intersect the $y$ unit length.
%cross section corresponding to the unit length.}

\section{Current phase relation of the S/$\mathrm{GdIr_2Si_2}$/S Josephson junction}
\label{sec:cpr}

\begin{figure}[t]
\includegraphics[width=0.85\columnwidth]{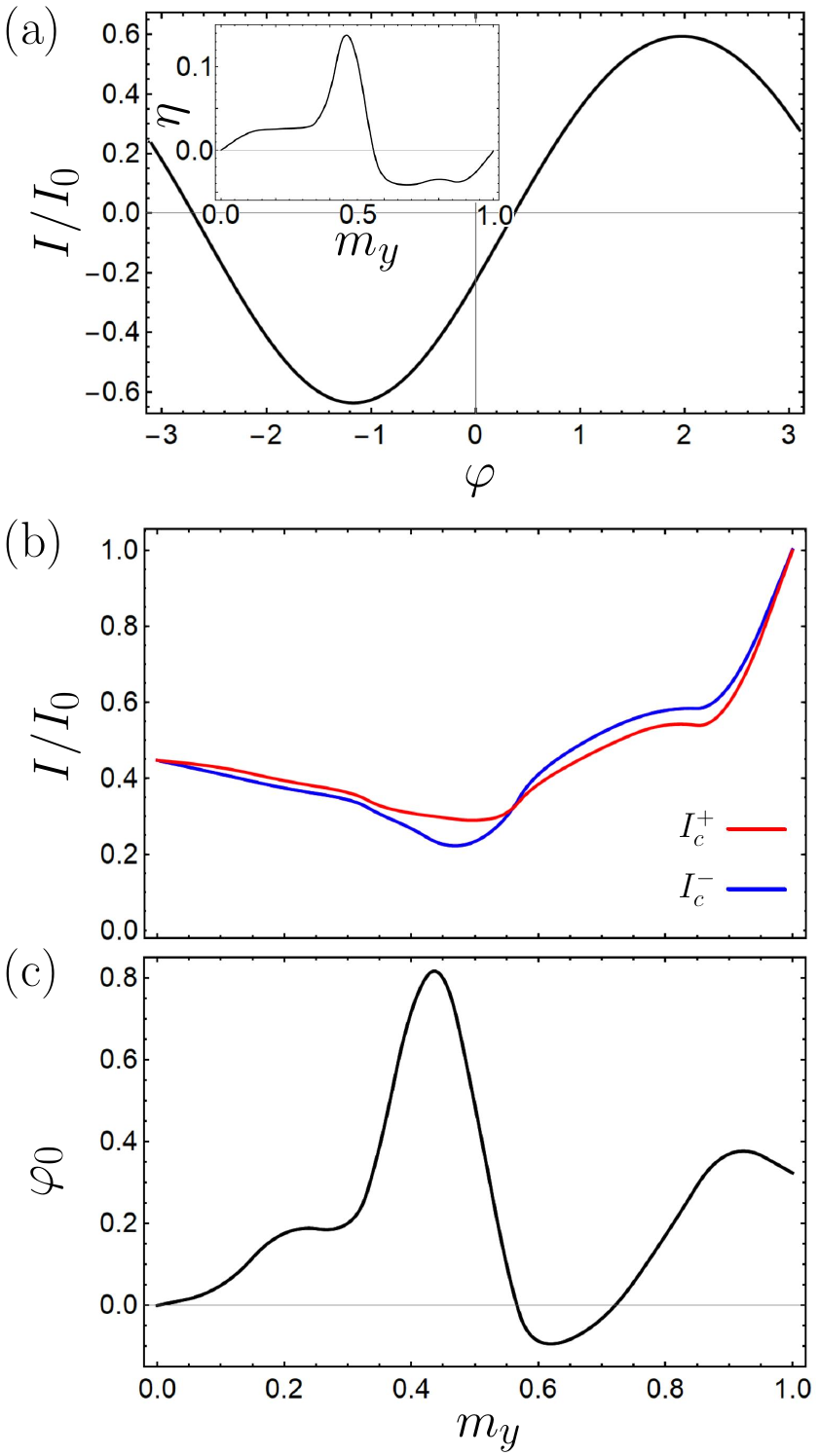}
\caption{Josephson characteristics of S/${\rm GdIr_2Si_2}$/S JJ with $L=30 a = 12.1$ nm. (a) CPR at $m_y = 0.9$. Inset: Josephson diode efficiency as a function of $m_y$. (b) Positive and negative critical Josephson currents as functions of $m_y$. (c) Anomalous ground state phase $\varphi_0$ as a function of $m_y$. The current is normalized to $I_0 = I_c^+(m_y=1)$.} 
 \label{fig:CPR_long}
\end{figure}

\begin{figure}[t]
\includegraphics[width=0.85\columnwidth]{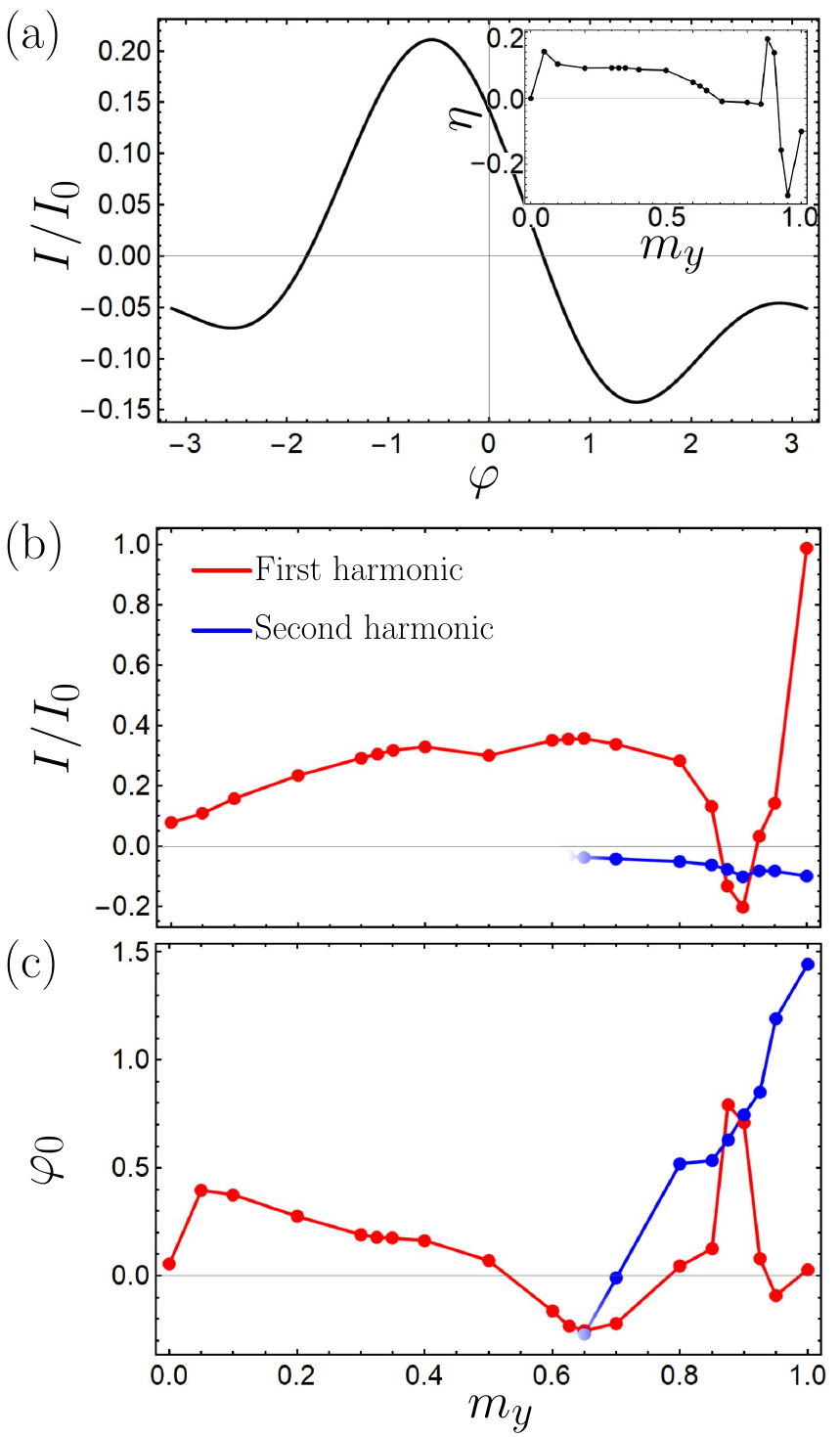}
\caption{Josephson characteristics of S/${\rm GdIr_2Si_2}$/S JJ with $L=20 a = 8.1$ nm. (a) CPR at $m_y = 0.875$. Inset: Josephson diode efficiency as a function of $m_y$. (b) Critical currents of the first and second harmonic contributions as functions of $m_y$. The second harmonic contribution becomes small and, therefore, is not shown in the figure at $m_y \lesssim 0.65$. (c) Anomalous ground state phases $\varphi_{0,1}$ (red) and $\varphi_{0,2}$ (blue) of the first and second harmonic contributions as functions of $m_y$. The other parameters are the same as in Fig.~\ref{fig:CPR_long}.} 
 \label{fig:CPR_short}
\end{figure}

The examples of CPRs calculated for two different lengths of the weak link region $L=30a = 12.1$~nm and $L=20a = 8.1$~nm are represented in Figs.~\ref{fig:CPR_long}(a) and \ref{fig:CPR_short}(a), respectively. For the longer JJ shown in Fig.~\ref{fig:CPR_long}, the CPR is nearly sinusoidal with small contribution of the higher harmonics. Nevertheless, it demonstrates the superconducting diode effect with small efficiency $\eta = (I_c^+ - I_c^-)/(I_c^+ + I_c^- ) < 0.15$, where $I_c^+  = {\rm max}[I(\varphi)]$ and $I_c^-  = -{\rm min}[I(\varphi)]$. As it was discussed above, in the considered case the magnet is an easy-plane magnet. Then, the magnetization is characterized by the only parameter $m_y$. Both critical currents $I_c^+ $ and $I_c^- $ are plotted as functions of $m_y$ in Fig.~\ref{fig:CPR_long}(b), and the corresponding diode efficiency is shown in the inset to Fig.~\ref{fig:CPR_long}(a). The Josephson current at $m_y<0$ is determined by the relation $I(-\varphi, -m_y) = -I(\varphi, m_y)$, which is dictated by the time reversal symmetry. 

Both critical currents exhibit a nonmonotonic dependence on $m_y$. Similarly, the diode efficiency as a function of $m_y$ is also highly nonmonotonic and even undergoes sign reversal.  This pronounced dependence of the critical current on the in-plane magnetization orientation arises from (i) the presence of multiple electronic bands crossing the Fermi level (in our case there are two spin-split bands) and (ii) the strong momentum dependence of the electronic structure near the Fermi level, which, in turn, is a direct consequence of the interplay between strong exchange and Rashba splittings. As evidenced by Fig.~\ref{fig:specific_phase}, the electronic spectra differ significantly depending on the momentum direction relative to the magnetization. The spectra along the magnetization direction [Fig.~\ref{fig:specific_phase}(a)], perpendicular to it [Fig.~\ref{fig:specific_phase}(b)], and along the diagonal directions [Figs.~\ref{fig:specific_phase}(c)-(d)] strongly differ in the vicinity of the Fermi surface. These differences give rise to a strongly anisotropic, spin-dependent Fermi velocity. 

Since the dominant contribution to the Josephson current comes from momentum directions close to the $x$-axis, the critical current is highly sensitive to the magnetization orientation. Specifically, the relevant electronic spectra at $m_y=0$ resemble those in Fig.~\ref{fig:specific_phase}(a), whereas at $m_y = 1$ they are similar to Fig.~\ref{fig:specific_phase}(b). Since the ${\rm GdIr_2Si_2}$ film is an easy-plane magnet, this direct link between the electronic structure and magnetization direction enables the adjustment of the diode efficiency through small rotations of the in-plane magnetization  thus enabling tunability of the diode effect via small variations of $m_y$. It is noteworthy that within the quasiclassical approximation for ferromagnetic materials with SOC, the critical current —including its higher harmonic components relevant for the Josephson diode effect— typically depends only on the magnitude of the interlayer magnetization and not on its direction relative to the current \cite{Buzdin2008,Bergeret2015,Konschelle2015}. In such models, reversing the magnetization ($m_y \to -m_y$) simply swaps the critical currents ($I_c^\pm  \to I_c^\mp $) and reverses the sign of the diode efficiency. A strong dependence of the critical current on the magnetization direction is predicted for junctions where the interlayer is 3D topological insulator with a Zeeman-split surface state \cite{Tanaka2009,Linder2010,Zyuzin2016,Rabinovich2020}; however, in those cases, the critical current is typically a monotonically increasing function of $m_y$, in contrast to the complex nonmonotonic behavior found here.

The anomalous phase shift $\varphi_0$ in the S/$\mathrm{GdIr_2Si_2}$/S junction also exhibits a highly nonmonotonic dependence on $m_y$. This behavior stands in stark contrast to the predictions of commonly used one-band model Hamiltonians, which typically yield a linear relationship $\varphi_0 \propto m_y$. In such simplified models, the torque exerted by the Josephson current on the magnetic moment of the ferromagnetic interlayer is governed by a constant coupling parameter $r$, defined via $\varphi_0 = r m_y$ \cite{Nashaat2019, Konschelle2009, Shukrinov2017, Guarcello2020}.
In our system, however, the nonmonotonic behavior of $\varphi_0(m_y)$ implies the existence of regions where the effective coupling $r(m_y) = d\varphi_0/dm_y$ changes sign. This suggests that the magnetization dynamics driven by the Josephson current will be significantly richer and more complex than in conventional scenarios. Understanding how this nontrivial $m_y$ dependence influences the magnetic response constitutes an important direction for future research and warrants a dedicated investigation.

\begin{figure}[t]
\includegraphics[width=0.85\columnwidth]{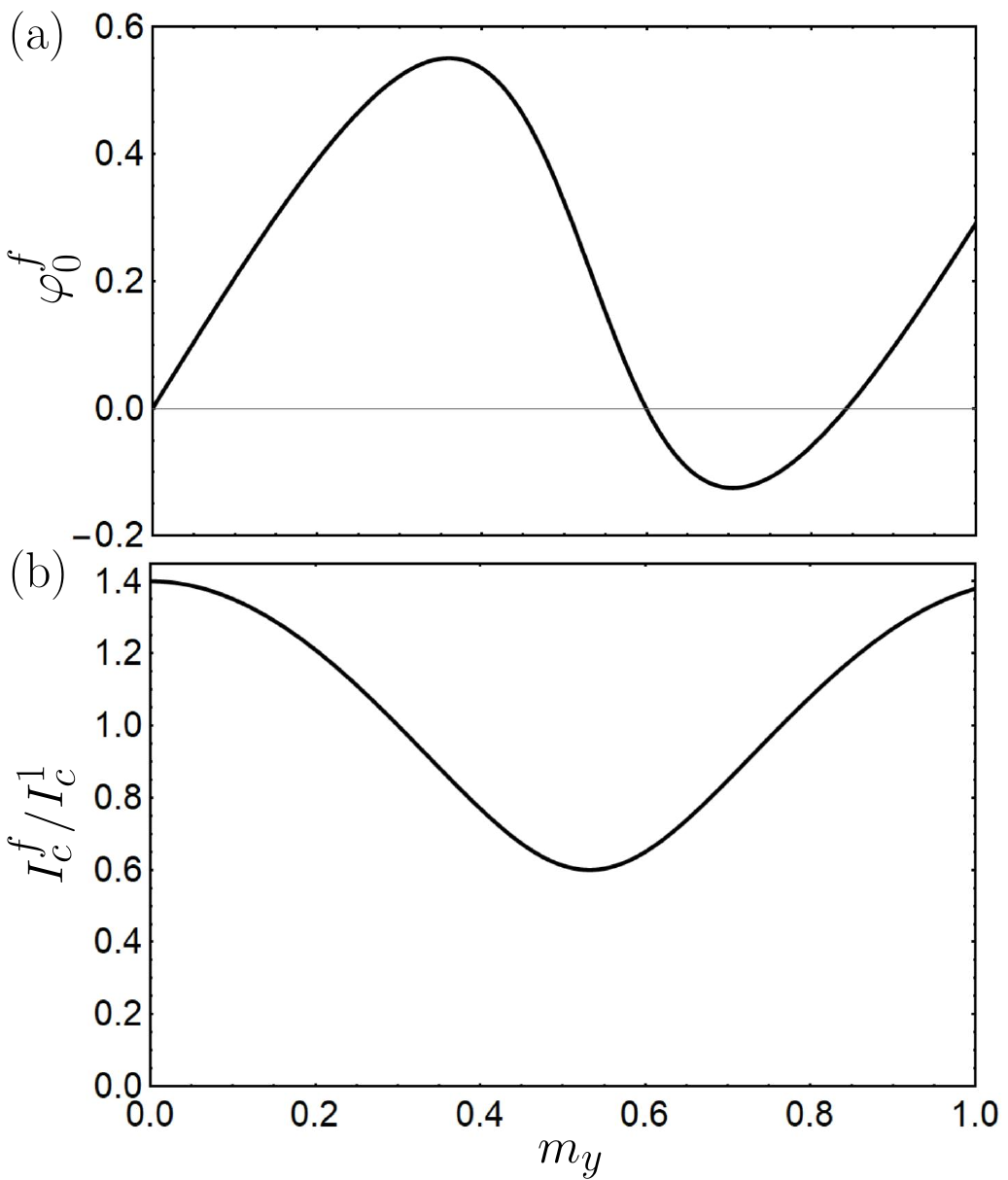}
\caption{Two-band minimal model for the Josephson current in a Rashba ferromagnet. (a) Anomalous phase shift $\varphi_0$ and (b) critical current $I_c^f$ versus $m_y$. Parameters: $I_c^2 = 0.4I_c^1$, $r^1 = 0.4$, $r^2 = 6.3$.} 
 \label{fig:model}
\end{figure}
The nonmonotonic dependence $\varphi_0(m_y)$ primarily arises from the contribution of multiple bands crossing the Fermi level. This assertion can be illustrated using a minimal model. Suppose that the Fermi level is crossed by two electronic bands, each exhibiting spin-momentum locking resulting from the Rashba-type spin splitting—similar to the case realized in the film under study. The current-phase relation for each individual band can be written as $I_{c}^i \sin(\varphi - r^i m_y)$, where $i = 1,2$ denotes the band index. The functional form $\varphi_0 \propto m_y$ is dictated by symmetry considerations. In general, $r^i$ may exhibit some dependence on $m_y$, which can be expanded as $r^i(m_y) = r^{i,0} + r^{i,2}m_y^2 + \dots$; however, in accordance with the literature \cite{Nashaat2019, Konschelle2009, Konschelle2015, Shukrinov2017, Guarcello2020} we restrict ourselves to a constant $r^i$. Likewise, $I_c^i$ may in principle depend on $m_y$—and indeed, for the real material this dependence is pronounced, as discussed above—but within this simple model we also treat $I_c^i$ as constant.

The total Josephson current is then given by $I = \sum_i I_{c}^i \sin(\varphi - r^i m_y)$. This expression can be recast in the form
\begin{align}
I = I_c^f \sin(\varphi - \varphi_0^f),
\label{eq:cpr_model1}
\end{align}
where
\begin{align}
&I_c^f = \sqrt{\left(\sum_i I_c^i \cos(r^i m_y)\right)^2 + \left(\sum_i I_c^i \sin(r^i m_y)\right)^2}, \label{eq:Ic_model1} \\ 
&\varphi_0^f = \arctan\left(\frac{\sum_i I_c^i \sin(r^i m_y)}{\sum_i I_c^i \cos(r^i m_y)}\right).
\label{eq:vaprhi_0_model1}
\end{align}

The resulting dependencies $\varphi_0(m_y)$ and $I_c^f(m_y)$ can have different shape depending on the choice of the model parameters $r^i$ and $I_c^i$, which are determined by the specific choice of the interlayer material. In particular, the nonmonotic behavior very similar to calculated above for our  S/$\mathrm{GdIr_2Si_2}$/S JJ is shown in Fig.~\ref{fig:model}. The results of this simple model clearly illustrate the physical origin of the nonmonotonic behavior of $\varphi_0(m_y)$ and $I_c^f(m_y)$.

A remark is in order here: for the S/$\mathrm{GdIr_2Si_2}$/S system considered in this work, it is not possible to isolate the contribution of each individual electronic band crossing the Fermi level in the chosen basis, because the TBH in Eq.~(\ref{eq:hamiltonian}) is not diagonal in the band index. However, we can effectively decouple the bands by artificially switching off the second band—i.e., setting $t^{(II),(II)} = t^{(I),(II)} = 0$. In this case, for the isolated band $I$ we indeed recover the linear dependence $\varphi_0 = r m_y$, with $r = 0.07$.

The CPR for the shorter junction, shown in Fig.~\ref{fig:CPR_short}(a), exhibits a clearly non-sinusoidal form. This shape is well-described by a two-harmonic fit, $I(\varphi) = I_{c1}\sin(\varphi - \varphi_{0,1}) + I_{c2}\sin(2 \varphi - \varphi_{0,2})$. The corresponding critical currents $I_{c1,2}$ and anomalous phase shifts $\varphi_{0,1,2}$ are presented in Figs.~\ref{fig:CPR_short}(b) and (c), respectively. For this junction, the anisotropy of the CPR parameters with respect to the in-plane magnetization orientation is even more pronounced. The second harmonic amplitude becomes significant only at larger values of $m_y$, and is therefore omitted from the plot at lower $m_y$. The diode efficiency again proves to be a sign-reversing function of $m_y$, with its magnitude increasing sharply at large $m_y$ in tandem with the growth of the second harmonic amplitude, which is consistent with general theories of the Josephson diode effect \cite{Baumgartner2022,Pal2022,Kim2024,Reinhardt2024,Sivakumar2025,Nadeem2023,Jeon2022,Jeon2026}.

\section{Discussion of other candidate materials}
\label{sec:discussion}
The GdIr$_2$Si$_2$ magnetic compound, which is the focus of the current study, belongs to the extensive $LnT_2X_2$ family, where $T$ is a transition metal and $X$ is a $p$-element from groups III-V \cite{Shatruk2019}. Most compounds within this family accommodate a ThCr$_2$Si$_2$-type (I-phase) crystal structure, while some, including GdIr$_2$Si$_2$, as well as other $Ln$Ir$_2$Si$_2$ compounds ($Ln$ = La, Ce, Pr, Nd, Sm) \cite{mihalik2011} and $Ln$Ni$_2$As$_2$ \cite{ElGhadraoui1988}, exhibit temperature-dependent polymorphism. In addition to the highly symmetric I-phase, they can crystallize in the CaBe$_2$Ge$_2$-type structure (space group P4/nmm).

The structural and magnetic properties of bulk $LnT_2X_2$ compounds are well-established \cite{Shatruk2019, Kliemt2020}. In these compounds, the selection of a lanthanide enables the control of the orientation of the $4f$ magnetic moments: they can be aligned in the $ab$ plane (as observed for Eu or Gd), along the $c$ axis (typical for Tb), or at a specific angle relative to the crystallographic axes (for instance, for Ho or Dy). 
Furthermore, by altering the transition metal atom in the compound, one can modify the strength of the spin-orbit interaction by selecting elements from the $3d$, $4d$, or $5d$ periods.

Nevertheless, in thin films that are several atomic layers thick, which are of interest for the creation of JJ interlayers, magnetic anisotropy as well as the preferred structure type may differ from those in the bulk materials. In these thin films, the desired balance between exchange and spin-orbit splitting can be effectively tuned by varying the $Ln$ and $T$ atoms. The change in $T$ within the period is also responsible for the adjustment of $E_\mathrm{F}$, which is further regulated by the element $X$. Thus, the compounds of the $LnT_2X_2$ family appear to be highly promising as potential candidates for the development of thin magnetic films with tailored properties.

The feasibility of experimentally producing the structures we put forward is an essential factor. The primary focus in this context is the potential to develop thin magnetic films, as their morphology will impact the physical characteristics of the weak link. Recently  the growth of thin films by molecular beam epitaxy (MBE) was demonstrated \cite{Prochaska2020}. These methods have been actively developing, demonstrating impressive success in growth of the ThCr$_2$Si$_2$-type thin films with enhanced crystallinity and surface smoothness \cite{Isceri2025}. 
Another important parameter of the suggested structure is critical magnetic temperature. The compounds with the tetragonal ThCr$_2$Si$_2$-type structure exhibit diverse critical magnetic temperatures (N\'eel or Curie), ranging from ultra-low, of few K, to over room temperature depending on the specific rare-earth or transition metal components \cite{Kliemt2020}. Regarding GdIr$_2$Si$_2$, the N\'eel temperature in the bulk crystal is 86 K. The $T_\mathrm{N}$ for thin films is unknown, yet it is likely to be not much lower. For example, in rare-earth based compound EuCd$_2$As$_2$, the $T_\mathrm{N}$ is 9.5 K in the bulk and 9.1 K in the thin film \cite{Wang2016}.

\section{Conclusions}
\label{sec:conclusions}

In this work, the current-phase relations of the Josephson junction through the $\mathrm{GdIr_2Si_2}$ thin-film interlayer have been investigated using a combination of DFT methods and Josephson current calculations in the formalism of the Bogoliubov-de Gennes equations. The electronic structure and magnetic properties of ultra-thin $\mathrm{GdIr_2Si_2}$ films that include two magnetic layers were thoroughly studied through the DFT calculations. It has been determined that the thin $\mathrm{GdIr_2Si_2}$ film of the stable I-phase exhibits easy-plane magnetism, characterized by the ferromagnetic alignment of magnetic moments within each of the two Gd layers, with the ferromagnetic interlayer arrangement being the most energetically advantageous. The effective tight-binding Hamiltonian, describing the bands of the $\mathrm{GdIr_2Si_2}$ film which cross the Fermi level, was constructed by fitting it to the DFT spectrum. 

Utilizing the constructed Hamiltonian, the Josephson current of the S/$\mathrm{GdIr_2Si_2}$/S Josephson junction was calculated numerically in the framework of the Bogoliubov-de Gennes approach. The current-phase relationships demonstrate a pronounced ground state anomalous phase shift of the order of unity and a pronounced Josephson diode effect with the diode efficiency $\lesssim 0.3$. The anomalous phase shift and the diode efficiency manifest strong anisotropy and sign reversing upon rotating the in-plane magnetization relative to the Josephson current direction, that makes this system promising for spintronics applications. The considered system has a high potential for further study. In particular, the electronic properties of the atomically thin $\mathrm{GdIr_2Si_2}$ films should be very sensitive to gating, what opens a way to electrical control of Josephson diode polarity and anomalous ground state phase shift. Another important step will be to investigate the magnetization dynamics and the switching of magnetization, which can be induced by the Josephson current in this system. Since the $\mathrm{GdIr_2Si_2}$ belongs to a wide family of magnetic compounds $LnT_2X_2$, variation in the composition of the thin films can provide a rich opportunity to control the electronic and magnetic properties in the interlayer of the Josephson junction.

\begin{acknowledgments}
The authors acknowledge Dmitry Usachev for his valuable advice on the choice of material. G.A.B. and I.V.B. acknowledge the support from Theoretical Physics and Mathematics Advancement Foundation “BASIS” via the project No. 23-1-1-51-1. The development of the effective tight-binding model was supported by the Russian Science Foundation via the RSF project No. 24-12-00152. The calculations of CPRs were supported by Grant from the ministry of science and higher education of the Russian Federation No. 075-15-2025-010. S.V.E. acknowledges support from the Government research assignment for ISPMS SB RAS, Project No. FWRW-2026-0008. I.A.Sh. gratefully acknowledges that DFT calculations were supported by the Ministry of Science and Higher Education of the
Russian Federation (State Task No. FSWM-2025-0009). E.V.C. acknowledges Saint-Petersburg State University for a research project 125022702939-2. This research was supported in part through computational resources of the supercomputer "SKIF Cyberia" of Tomsk State University.
\end{acknowledgments}

\appendix

\section*{Appendix: Sensitivity of the CPR to superconducting and interface parameters and temperature}

\begin{figure}[t]
\includegraphics[width=0.85\columnwidth]{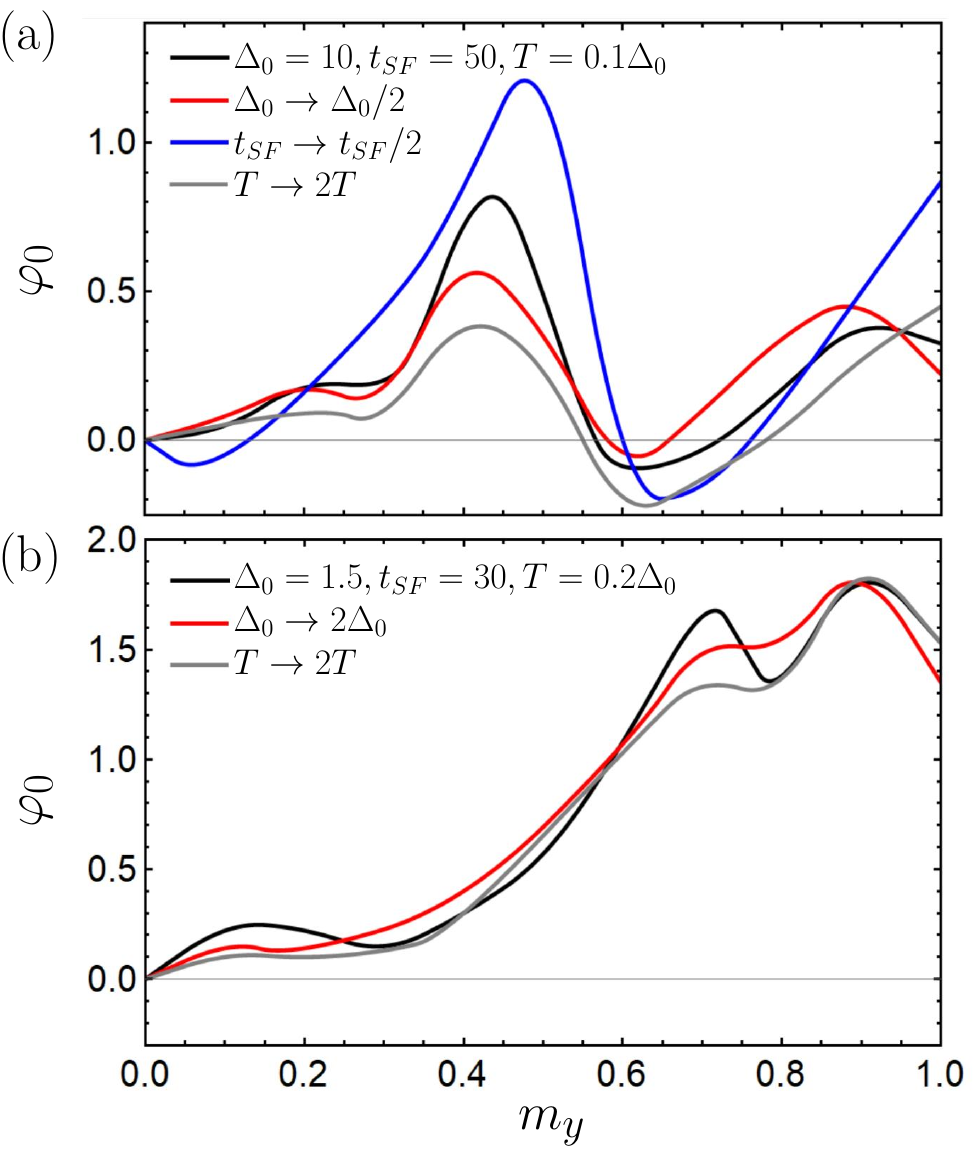}
\caption{The black solid curve in panel (a) shows the anomalous phase shift $\varphi_0(m_y)$ for the original parameter set corresponding to Fig.~\ref{fig:CPR_long}(c). The remaining curves in panels (a) and (b) illustrate how this dependence is modified under variations of the environmental parameters—specifically, the superconducting order parameter $\Delta_0$, the S/F interface hopping amplitude $t_{SF}$, and the temperature $T$.} 
 \label{fig:varphi_0_parameters}
\end{figure}

\begin{figure}[t]
\includegraphics[width=0.85\columnwidth]{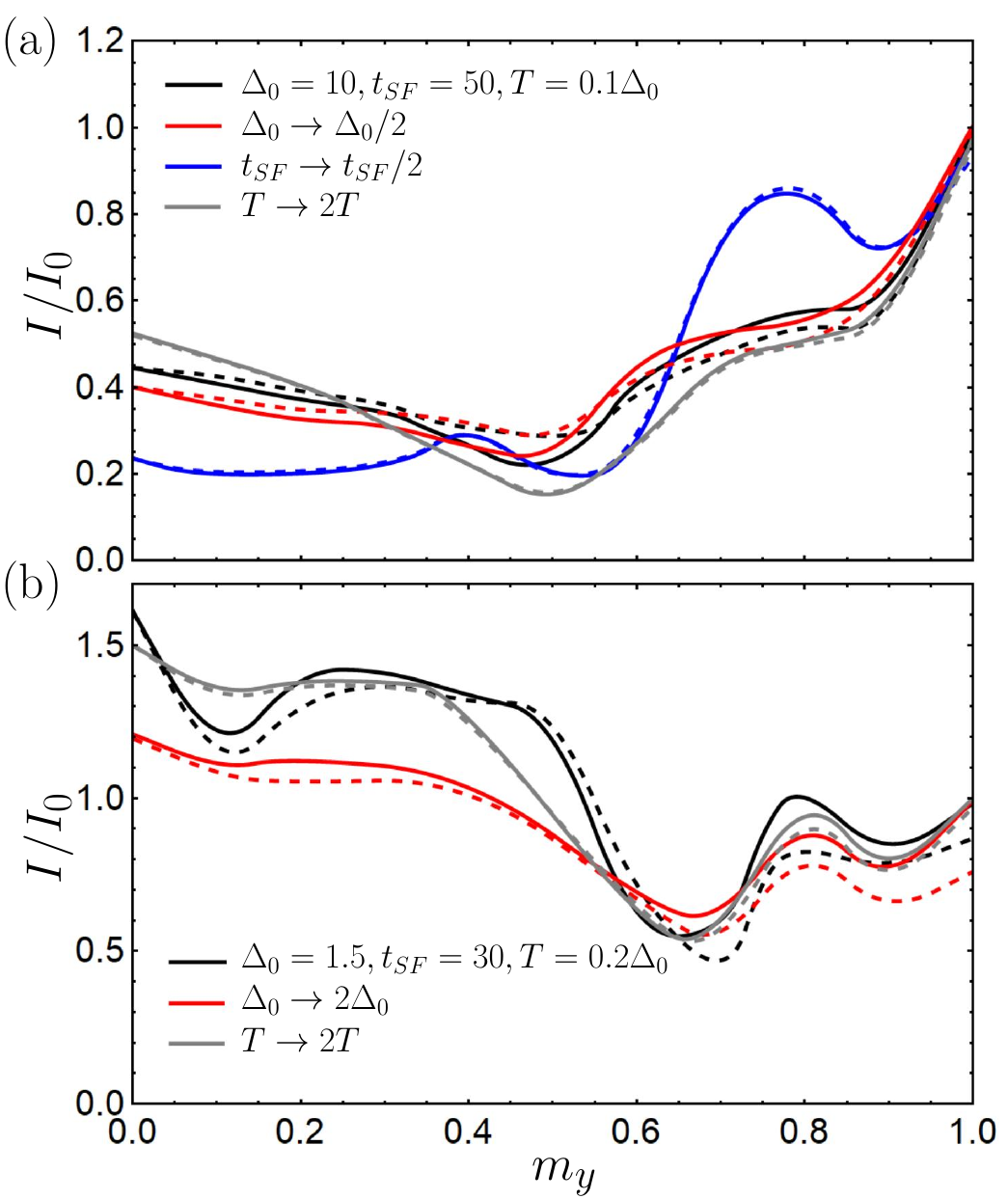}
\caption{The black curves in panel (a) show the critical currents $I_c^+ (m_y)$ (solid) and $I_c^- (m_y)$ (dashed) for the original parameter set corresponding to Fig.~\ref{fig:CPR_long}(b). The remaining curves in panels (a) and (b) illustrate how this dependence is modified under variations of the environmental parameters—specifically, the superconducting order parameter $\Delta_0$, the S/F interface hopping amplitude $t_{SF}$, and the temperature $T$. The current is normalized to $I_0=I_c^+(m_y=1)$, in case (a) $I_0=130$~A/m, (b) $I_0=6$~A/m.} 
 \label{fig:critical_parameters}
\end{figure}

The observed sensitivity of the anomalous phase shift $\varphi_0$ and the critical current $I_c$ to variations in the superconducting order parameter and the interface hopping amplitude $t_{SF}$, as shown in Figs.~\ref{fig:varphi_0_parameters} and \ref{fig:critical_parameters}, arises from the contribution of multiple bands crossing the Fermi level and/or the presence of higher harmonics in the current-phase relation. Indeed, if each band contributed only via the first harmonic, as assumed in Eq.~(\ref{eq:cpr_model1}), and if the critical currents of both bands exhibited the same functional dependence on $\Delta_0$ and $t_{SF}$, then $\varphi_0$ and the normalized critical current $I_c^f(m_y)/I_c^f(m_y=1)$ would be completely independent of these parameters. In such a scenario, $\Delta_0$ and $t_{SF}$ would merely enter as a global amplitude factor multiplying the individual band contributions $I_c^{i}$, leaving the functional form of the CPR unaffected.

In reality, however, even when higher-harmonic contributions to the Josephson current are small—as is the case for the results shown in Fig.~\ref{fig:CPR_long}—the functional dependence of $I_c^i$ on $t_{SF}$ differs between bands. This is because the electronic band dispersions, and consequently the Fermi surface geometries, are distinct for each band. As a result, despite the same nominal interface hopping amplitude, the effective transparency of the S/F interface and the magnitude of the proximity-induced superconducting gap in the $\mathrm{GdIr_2Si_2}$ region beneath the superconducting electrodes can differ between bands. Consequently, according to Eqs.~(\ref{eq:cpr_model1})-(\ref{eq:vaprhi_0_model1}), the dependence on $\Delta_0$ and $t_{SF}$ does not cancel out in the anomalous phase shift $\varphi_0$ or in the normalized critical current $I_c(m_y)/I_c(m_y=1)$, as clearly demonstrated by the results presented in Figs.~\ref{fig:varphi_0_parameters} and \ref{fig:critical_parameters}.

However, despite the quantitative differences, the data presented in the figures demonstrate that all key qualitative features remain robust and are governed solely by the properties of the electronic band spectrum near the Fermi level—namely, the number of bands crossing the Fermi level and their dispersion characteristics. In other words, these essential features are determined exclusively by the intrinsic material properties of the interlayer.

Figs.~\ref{fig:varphi_0_parameters} and \ref{fig:critical_parameters} also include the results of a sensitivity analysis of $\varphi_0(m_y)$ and $I_c(m_y)$ with respect to temperature variations. The results clearly demonstrate that temperature changes do not qualitatively alter these dependencies.

\begin{figure}[t]
\includegraphics[width=0.85\columnwidth]{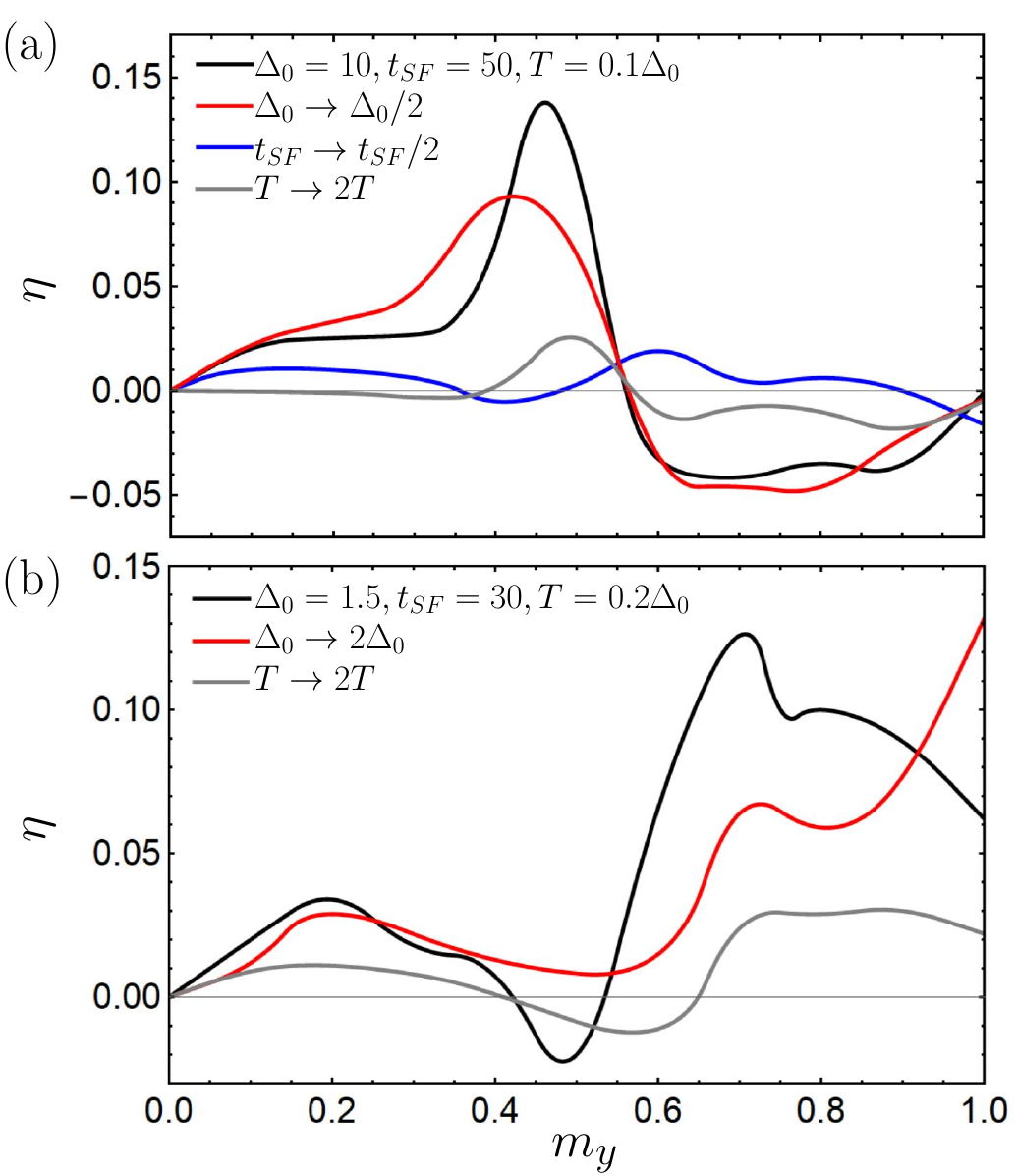}
\caption{The black curve in panel (a) shows the Josephson diode efficiency $\eta(m_y)$ for the original parameter set corresponding to Fig.~\ref{fig:CPR_long}(a). The remaining curves in panels (a) and (b) illustrate how this dependence is modified under variations of the environmental parameters—specifically, the superconducting order parameter $\Delta_0$, the S/F interface hopping amplitude $t_{SF}$, and the temperature $T$.} 
 \label{fig:eta_parameters}
\end{figure}

As discussed in the main text, the only junction characteristic that exhibits a pronounced dependence on $\Delta_0$, $t_{SF}$, and temperature is the magnitude of the diode effect. This sensitivity arises because the diode efficiency is governed by the amplitude of the second harmonic of the Josephson current. Within our minimal two-band model, the diode efficiency can be expressed as:
\begin{align}
\eta = \frac{I_{c2} \sin(2\varphi_{01} - \varphi_{02})}{I_{c1}},
\label{eq:eta_model}
\end{align}
where $I_{c1}$ and $I_{c2}$ are the critical currents of the first and second harmonics, respectively, computed using Eq.~(\ref{eq:cpr_model1})-(\ref{eq:vaprhi_0_model1}), applied separately to the first- and second-harmonic contributions.

It is well established in the literature that the amplitudes of the first and second harmonic components of the critical current exhibit distinct dependencies on $\Delta_0$ and the interface hopping parameter. While the exact form of this dependence may vary across different junction types \cite{Golubov2004}, a general feature is that $I_{c,2}^i / I_{c,1}^i \to 0$ as $T \to T_c$ or $\Delta_0 \to 0$—i.e., the current-phase relation becomes purely sinusoidal as superconductivity weakens. The same behavior holds in the limit $t_{SF} \to 0$. Owing to the different functional dependencies of $I_{c1}$ and $I_{c2}$, the diode efficiency $\eta$ exhibits a pronounced sensitivity to variations in $\Delta_0$ and $t_{SF}$.

The calculated diode efficiency $\eta$ for the S/$\mathrm{GdIr_2Si_2}$/S junction under study—obtained without invoking the minimal model—is presented in Fig.~\ref{fig:eta_parameters} and indeed demonstrates a strong sensitivity to changes in $\Delta_0$ and $t_{SF}$. However, it is important to emphasize that, in contrast to single-band models, $\eta$ in our system is not proportional to $\sin(r m_y)$; rather, it exhibits a sign-changing behavior as a function of $m_y$ over the interval $m_y \in (0,1)$. This property stems from the two-band nature of Josephson transport in the system under consideration and remains robust against variations in the superconducting environment and interface parameters.

\bibliography{GdIr2Si2}

\end{document}